\documentclass[11pt,twoside]{article}
\usepackage{asp2014}
\pdfoutput=1

\aspSuppressVolSlug
\resetcounters

\begin{document}

\title{Adding Workflow Management Flexibility to LSST Pipelines Execution}

\author{Michelle~Gower,$^1$ Mikolaj Kowalik,$^1$ Nate~Lust,$^2$ James Bosch,$^2$\\ and Tim Jenness$^3$}
\affil{$^1$National Center for Supercomputing Applications, University of Illinois at Urbana-Champaign, Urbana, IL 61801, USA \email{mgower@illinois.edu}}
\affil{$^2$Department of Astrophysical Sciences, Princeton University, Princeton, NJ 08544, USA}
\affil{$^3$Rubin Observatory Project Office, Tucson, AZ 85719, USA}

\paperauthor{Michelle~Gower}{mgower@illinois.edu}{0000-0001-9513-6987}{University of Illinois Urbana-Champaign}{National Center for Supercomputing Applications}{Urbana}{IL}{61801}{USA}
\paperauthor{Mikolaj~Kowalik}{mxk@illinois.edu}{000-0002-9801-5969}{University of Illinois Urbana-Champaign}{National Center for Supercomputing Applications}{Urbana}{IL}{61801}{USA}
\paperauthor{Nate~B.~Lust}{nlust@astro.princeton.edu}{0000-0002-4122-9384}{Princeton University}{Department of Astrophysical Sciences}{Princeton}{NJ}{08544}{USA}
\paperauthor{James~F.~Bosch}{jbosch@astro.princeton.edu}{0000-0003-2759-5764}{Princeton University}{Department of Astrophysical Sciences}{Princeton}{NJ}{08544}{USA}
\paperauthor{Tim~Jenness}{TJenness@lsst.org}{0000-0001-5982-167X}{Rubin Observatory Project Office}{950 N. Cherry Ave.}{Tucson}{AZ}{85719}{USA}




\begin{abstract}

Data processing pipelines need to be executed at scales ranging from
small runs up through large production data release runs resulting in
millions of data products.  As part of the Rubin Observatory's pipeline
execution system, BPS is the abstraction layer that provides an interface
to different Workflow Management Systems (WMS) such as HTCondor and PanDA.
During the submission process, the pipeline execution system interacts
with the Data Butler to produce a science-oriented execution graph from
algorithmic tasks.  BPS converts this execution graph to a workflow graph
and then uses a WMS-specific plugin to submit and manage the workflow.
Here we will discuss the architectural design of this interface and
report briefly on the recent production of the Data Preview 0.2 release
and how the system is used by pipeline developers.

\end{abstract}

\section{Introduction}

A large number of images will be taken during the 10-year Legacy Survey
of Space and Time \citep[LSST;][]{2019ApJ...873..111I, I08_adassxxxii}.
Processing of this data will be done at scales ranging from small runs
up through large production data release runs.  Data processing will be
done using data management and execution systems for the Rubin Observatory
LSST Science Pipelines with the Data Butler and science-oriented execution
graphs \citep{C24_adassxxxii,2022SPIE12189E..11J}.

The Data Butler is the system that abstracts the data ac\-cess de\-tails from
the pipe\-line de\-vel\-opers.  A Quantum is the work to be done on a single set
of inputs (e.g., remove instrument signature on the raw image for detector
3 of exposure 12345). These Quanta take inputs and produce outputs which
affect the order in which they can be executed.  This science-oriented
execution directed acyclic graph is called a QuantumGraph.

This QuantumGraph is executable by LSST middleware on a single machine.
At this point, the graph has no runtime information (such as command
lines or required memory) that would be needed to run jobs via
a workload management system (e.g., Slurm).  BPS is the middleware
that converts the Quantum Graph into a workflow by adding the runtime
information. It provides a layer of abstraction so different WMS can be
used with minimal user-facing changes in either commands or configuration
as it was quickly apparent that LSST needs to have the flexibility to
use different WMS. For example, users at different sites want to use
the WMS they are familiar with or the release processing would like to
try a new WMS.

\section{Submission}
Users interact with BPS using the same middleware terminology as used
in notebooks and single machine execution to define the input data
and pipeline.  The submission process is broken up into phases:

\begin{enumerate}
\item \textit{acquire}: Create the QuantumGraph or read an existing one.

\item \textit{cluster}: Group Quanta into clusters for efficiency (see
subsection \ref{clustering} for more details).

\item \textit{transform}: Create a generic workflow graph from the
clustered QuantumGraph (see \S\ref{generic_workflow} for more
details).

\item \textit{prepare}: Convert the generic workflow graph into the
representation required by the workflow management system.  Being able
to save this representation to disk is especially helpful for debugging
submission issues.

\item \textit{submit}: Finally submit the workflow representation for
execution.
\end{enumerate}

\subsection{Clustering} \label{clustering}

Early running of pipelines showed that many of the Quanta were
executed very quickly, for example in less than a minute.  The job
overhead would be too large for efficient running of those quick jobs
and many batch schedulers do not handle well thousands of minute long
jobs.  We added an option for the BPS submission process to
group Quanta into clusters and then each cluster corresponds to
a compute job.

The disadvantages of clustering is that the clusters are black
boxes to the workflow and batch services.  This means that workflow
and batch services cannot monitor or retry at the Quantum level.
The LSST execution middleware can skip Quanta that have already
successfully run, and more work is planned on Quantum-level reporting
and real-time logging.  The bigger disadvantage is that workflows
generally stop execution downstream of a failed job.  This means
a single Quantum failure could keep many other Quanta from running
that could actually be run just because of the way they were
clustered.

The method used to perform the clustering is a submit time configuration.
Clustering is independent of the workflow management system.  There
are currently two clustering methods supported: the
original behavior where each Quantum is its own cluster and a user-specified
clustering.  The user-specified clustering is done by common data values
(e.g., same detector) to help maximize the execution of Quanta in the
workflow.

\subsection{Generic Workflow} \label{generic_workflow}
The generic workflow is a directed a\-cyclic graph con\-tain\-ing run\-time
in\-form\-a\-tion need\-ed to sub\-mit to a workflow management system.  Each
cluster becomes a job in the workflow.  The job's command line and
resource requirements (e.g., memory, cpu, walltime, batch queue) are
added here.  In addition to jobs representing clusters, two special jobs
are added to the workflow.  There is an initial job that saves information
about the run such as software stack versions and configuration.  Some
of this information is used by later jobs.  The other special job is the
last job in the workflow.  The final job records the workflow output
datasets in the Butler repository.  This always needs to be executed
whether there was a job failure or not.  Because failures halt execution
along paths in the workflow, this final job cannot be added as a normal
job to the workflow graph.

\section{After Submission}
There are a few actions users want to take on submitted workflows.  The
ones implemented so far include \texttt{report}, \texttt{cancel}, and \texttt{restart}.
\texttt{report} gets the workflow and job status and summarize
the results using LSST labels.  Internally this uses the WMS plugin to
ask for the statuses.
\texttt{cancel} aborts a submitted workflow killing any running
jobs.  The final job for the canceled workflow should still run to ensure
the consistency of the central repository state.  Internally this uses
the WMS plugin to abort the workflow and its jobs.
\texttt{restart} has two modes.
As-is restart can be used if a workflow died due to infrastructure
issues to retry the failed jobs as is and continue the workflow.
This requires support from the workflow system.
Re-submission could add a couple of options to the BPS configuration
as well as other changes, including software version, and submit again
which would produce a new QuantumGraph without the Quanta successfully
completed.  This does not require restart support from the workflow
system.

\section{Notable Workflow Needs}
There have been a few workflow features that LSST workflows need that
cause difficulties for various workflow management systems.

The Butler abstracts the data location from the code.  It can directly
read and write files in the repository making local cached copies
when needed.  Therefore, the science inputs and outputs are currently
not included in the workflow, and the workflow system can expect the
data to be in place when the jobs run.  This means that the workflow
system needs to support job dependencies without data dependencies or
the plugin will need to fake the data dependencies.

There is the final job (see subsection \ref{generic_workflow}) that should be
executed regardless of whether a job failed.  This is not typical behavior
for jobs in most workflow systems, so the workflow system needs to have
special support for this kind of job.

In order to efficiently share compute resources, many workload schedulers
require knowing how much memory a job will need.  Different input data can
cause a particular job to require more memory than expected especially
during development.  Having the job automatically retry with a larger
memory request if it was killed for memory is a desired feature.

\section{Existing Plugins and Future Work}
BPS was primarily constructed using HTCondor's \citep{10.1002/cpe.938}
DAGMan on a native HTCondor pool at NCSA.  During the project's
construction period, pipeline developers also used this to run tests.
Mini processing campaigns were also run every few weeks as larger
scale tests.

Data Preview 0.2 \citep[DP0.2;][]{RTN-039}, was generated by
executing the workflows via a BPS plugin for the PanDA workflow system
\citep{10.1088/1742-6596/331/7/072024}.  This processing was done on
the Interim Data Facility in a Google Cloud environment \citep{2021arXiv211115030O}.
While not currently officially maintained and supported by the core team,
there is also a Pegasus \citep{10.1016/j.future.2014.10.008} plugin and
a Parsl \citep{10.1145/3307681.3325400} plugin.

Development is ongoing across the entire middleware.  Besides updating
for these changes and making plugin improvements, there are a few areas
where work is planned.  While still overall useful, BPS's report has
became less informative for the user with clustering.  They want to know
which Quanta failed and for what reason.  Clustering also has caused
problems with the current command lines.  Work is ongoing for a special
job runner that will solve the long command line issue as well as open
avenues for other job-related functions across multiple workflow
management systems such as sending output to real-time log aggregators.

\acknowledgements We thank Richard Dubois for his review of this
manuscript.  This material is based upon work supported in part by the
National Science Foundation through Cooperative Agreement AST-1258333
and Cooperative Support Agreement AST-1202910 managed by the Association
of Universities for Research in Astronomy (AURA), and the Department
of Energy under Contract No.\ DE-AC02-76SF00515 with the SLAC National
Accelerator Laboratory managed by Stanford University. Additional Rubin
Observatory funding comes from private donations, grants to universities,
and in-kind support from LSSTC Institutional Members.



\begin{thebibliography}{}
    \expandafter\ifx\csname natexlab\endcsname\relax\def\natexlab#1{#1}\fi
    \expandafter\ifx\csname url\endcsname\relax
      \def\url#1{\texttt{#1}}\fi
    \expandafter\ifx\csname urlprefix\endcsname\relax\def\urlprefix{URL }\fi
    \providecommand{\eprint}[2][]{\url{#2}}

    \bibitem[{Babuji et~al.(2019)}]{10.1145/3307681.3325400}
    Babuji, Y., et~al. 2019, in Proceedings of the 28th International Symposium on
      High-Performance Parallel and Distributed Computing (New York, NY, USA:
      Association for Computing Machinery), HPDC '19, 25–36.
      doi:10.1145/3307681.3325400

    \bibitem[{Deelman et~al.(2015)}]{10.1016/j.future.2014.10.008}
    Deelman, E., et~al. 2015, Future Generation Computer Systems, 46, 17.
      doi:10.1016/j.future.2014.10.008

    \bibitem[{{Ivezi{\'c}} et~al.(2019)}]{2019ApJ...873..111I}
    {Ivezi{\'c}}, {\v Z}., et~al. 2019, \apj, 873, 111.
      doi:10.3847/1538-4357/ab042c, arXiv:0805.2366

    \bibitem[{{Jenness} et~al.(2022)}]{2022SPIE12189E..11J}
    {Jenness}, T., et~al. 2022, in Software and Cyberinfrastructure for Astronomy
      VII, vol. 12189 of Proc.\ SPIE, 1218911. doi:10.1117/12.2629569,
      arXiv:2206.14941

    \bibitem[{{Lust} et~al.(2023)}]{C24_adassxxxii}
    {Lust}, N., et~al. 2023, in ADASS XXXII, edited by S.~{Gaudet}, S.~{Gwyn},
      P.~{Dowler}, D.~{Bohlender}, \& A.~{Hincks} (San Francisco), vol. TBD of ASP
      Conf. Ser., 999 TBD

    \bibitem[{Maeno et~al.(2011)}]{10.1088/1742-6596/331/7/072024}
    Maeno, T., et~al. 2011, Journal of Physics: Conference Series, 331, 072024.
      doi:10.1088/1742-6596/331/7/072024

    \bibitem[{{O'Mullane} et~al.(2021)}]{2021arXiv211115030O}
    {O'Mullane}, W., et~al. 2021, in ADASS XXI, ASP Conf.\ Ser., in press.
      arXiv:2111.15030

    \bibitem[{{O'Mullane} et~al.(2023)}]{I08_adassxxxii}
    --- 2023, in ADASS XXXII, edited by S.~{Gaudet}, S.~{Gwyn}, P.~{Dowler},
      D.~{Bohlender}, \& A.~{Hincks} (San Francisco), vol. TBD of ASP Conf. Ser.,
      999 TBD. arXiv:2211.13611

    \bibitem[{Thain et~al.(2005)Thain, Tannenbaum, \& Livny}]{10.1002/cpe.938}
    Thain, D., Tannenbaum, T., \& Livny, M. 2005, Concurrency and computation:
      practice and experience, 17, 323. doi:10.1002/cpe.938

    \bibitem[{Yanny et~al.(2022)Yanny, Kuropatkin, Lin, Adelman-McCarthy, Slater,
      \& Chiang}]{RTN-039}
    Yanny, B., Kuropatkin, N., Lin, H., Adelman-McCarthy, J., Slater, C., \&
      Chiang, H.-F. 2022, {Compute Resource Usage of DP0.2 Production Run}.
      \urlprefix\url{https://rtn-039.lsst.io/}

    \end{thebibliography}
\end{document}